\documentclass[fleqn,useAMS,usenatbib]{mn2e}
\usepackage[english]{babel}

\usepackage{fancyhdr}
\usepackage{amsfonts}
\usepackage{amsmath}
\usepackage{amssymb}
\usepackage{multicol}
\usepackage{layout}
\usepackage{graphicx}
\usepackage{epsfig}
\usepackage{verbatim}
\usepackage[T1]{fontenc}
 \usepackage{aecompl}
\usepackage{color}
\usepackage{multirow}

\usepackage{times}
\usepackage{natbib}

\newif\ifAMStwofonts
\AMStwofontstrue

\graphicspath{{./fig/}}

\title{How clustering dark energy affects matter perturbations}

\author[Mehrabi et~al.]{A. Mehrabi$^1$\thanks{mehrabi@basu.ac.ir}, 
S. Basilakos$^2$\thanks{svasil@academyofathens.gr} and F. Pace$^{3}$\\
$^1$ Department of Physics, Bu-Ali Sina University, Hamedan 65178, Iran \\
$^2$ Academy of Athens, Research Center for Astronomy \& Applied
  Mathematics, Soranou Efessiou 4, 11-527, Athens, Greece \\
$^3$ Jodrell Bank Centre for Astrophysics, School of Physics and Astronomy, The University of Manchester, Manchester, 
M13~9PL, UK}
\date{Accepted ?, Received ?; in original form \today}

\begin{document}
\maketitle

\label{firstpage}

\begin{abstract}
The rate of structure formation in the Universe is different in homogeneous and clustered dark energy models. 
The degree of dark energy clustering depends on the magnitude of its 
effective sound speed $c^{2}_{\rm eff}$ and for 
$c_{\rm eff}=0$ dark energy clusters in a similar 
fashion to dark matter while for $c_{\rm eff}=1$ it stays 
(approximately) homogeneous. 
In this paper we consider two distinct equations of state for the dark energy component, $w_{\rm d}=const$ and 
$w_{\rm d}=w_0+w_1\left(\frac{z}{1+z}\right)$ with $c_{\rm eff}$ as a free parameter and we try to constrain the 
dark energy effective sound speed using current available data including SnIa, Baryon Acoustic Oscillation, CMB shift 
parameter ({\em Planck} and {\em WMAP}), Hubble parameter, Big Bang Nucleosynthesis and the growth rate of structures 
$f\sigma_{8}(z)$. At first we derive the most general form of the equations governing dark matter and dark energy 
clustering under the assumption that $c_{\rm eff}=const$. 
 Finally, performing an overall likelihood analysis we find
that the likelihood function peaks at $c_{\rm eff}=0$, however  
the dark energy sound speed 
is degenerate with respect to the cosmological parameters, namely $\Omega_{\rm m}$ and $w_{\rm d}$.

\end{abstract}


\section{Introduction}\label{intro}
We are living in a special epoch of the cosmic history where the expansion of the Universe is accelerated due to 
an unknown energy component, usually dubbed dark energy (DE). This acceleration has been discovered observationally 
using the luminosity distance of Type Ia supernovae (SnIa) 
\citep{Perlmutter:1996ds,Perlmutter:1997zf,Perlmutter:1998np,Riess:2004nr,Astier:2005qq,Jha:2006fm}. In addition to 
this, other observations including the cosmic microwave background (CMB) 
\citep{Bennett:2003bz,Spergel:2003cb,Spergel:2006hy,Planck:2015xua,Ade:2015yua}, large scale structures (LSS) 
\citep{Hawkins:2002sg,Tegmark:2003ud,Cole:2005sx} and baryon acoustic oscillation (BAO) 
\citep{Eisenstein:2005su,Seo:2005ys,Blake:2011en} support an accelerated expansion. At a fundamental level 
there are two different approaches to describe the phenomenon of the cosmic acceleration and indeed many efforts are 
devoted to investigate its deep nature both observationally and theoretically. One way is to consider a fluid with a 
sufficiently negative pressure dubbed DE and the other is based on the modification of the laws of gravity on large 
scales. The first approach comes in many different scenarios. The simplest one is a very tiny cosmological constant 
$\Lambda$ in Einstein field equations that has a (negative) pressure equal to its energy density and equation-of-state 
parameter $w_{\rm d}=\frac{p_{\rm d}}{\rho_{\rm d}}=-1$ \citep{Weinberg:1989,Sahni:1999gb,Peebles:2002gy}. The overall 
theoretical cosmological model (cosmological constant plus cold dark matter to explain galaxy rotation curves and the 
potential well for structure formation) is called $\Lambda$CDM model. 
Despite being highly consistent with observational data, the $\Lambda$CDM model suffers of two theoretical problems, 
namely the fine-tuning and the cosmic coincidence problem \citep{Weinberg:1989,Sahni:1999gb,Peebles:2002gy}. 
Differently from the cosmological constant case with equation of state (EoS) $w_{\rm d}=-1$, other dynamical models 
have been largely studied in the literature and usually categorised in two branches, quintessence models 
\citep{ArmendarizPicon:2000dh,Copeland:2006wr} and k-essence models 
\citep{ArmendarizPicon:1999rj,ArmendarizPicon:2000dh,Chiba:1999ka,Chiba:2009nh,Amendola:2010}.

The simplest way to modify gravity is to consider Einstein-Hilbert Lagrangian as a generic function of the Ricci 
scalar $R$ \citep[$f(R)$ theories,][]{Schmidt:1990gb,Magnano:1993bd,Dobado:1994qp,Capozziello:2003tk,Carroll:2003wy}
or add extra-dimension models like in the DGP model \citep*{Dvali:2000hr}. 
Understanding which class of models is the real one is one of the biggest challenges for cosmology.  

In addition to the background evolution, large scale structures provide valuable information about the nature of dark 
energy \citep{Tegmark:2003ud,Tegmark:2006az}. Primordial matter perturbations grow throughout the cosmic history 
and their growth rate depends on the overall energy budget and on the properties of the cosmic fluids. DE slows down 
the growth rate of large-scale structures. Structures grow due to gravitational instability and DE acts opposing and 
reducing the growth rate. The growth rate of structures can be measured from the redshift space distortion (RSD). 
Inward peculiar velocities of large-scale structures generate a distortion that is directly related to the matter 
density contrast.

Since the cosmological constant does not change in space and time, it can not cluster like dark matter (DM) and it 
has a negligible contribution to the energy density of the universe at high redshift. On the other hand, dynamical DE 
can cluster and the amount of clustering depends strongly on its effective sound speed. The effective sound speed is 
defined as $c_{\rm eff}^2=c_{\rm e}=\frac{\delta p_{\rm d}}{\delta\rho_{\rm d}}$ 
(hereafter we use $c_{\rm e}$) where $\delta p_{\rm d}$ and $\delta\rho_{\rm d}$ are the pressure and energy density 
perturbations for DE respectively and coincides with the actual sound speed in the dark energy comoving rest frame 
\citep{Hu:1998kj}. In quintessence models we have $c_{\rm e}\simeq 1$ so DE perturbations can not grow on sub-horizon 
scales while in k-essence models the effective sound speed can be tiny ($c_{\rm e}\ll 1$) 
\citep{Garriga:1999vw,ArmendarizPicon:1999rj,ArmendarizPicon:2000dh,Babichev:2006vx,Akhoury:2011hr} and DE 
perturbations grow similarly to dark matter (DM) perturbations. The possibility of DE clustering has been studied by 
many authors \citep{Erickson:2001bq,Bean:2003fb,Hu:2004yd,Ballesteros:2008qk,dePutter:2010vy,Sapone:2012nh,
Batista:2013oca,Dossett:2013npa,Basse:2013zua,Batista:2014uoa,Pace:2014taa,Steigerwald:2014ava}. 
In particular, it has been shown that the homogeneous DE scenario fails to reproduce the observed concentration 
parameter of the massive galaxy clusters \citep{Basilakos:2009mz}. In this framework, \cite{dePutter:2010vy} pointed 
out that CMB and LSS slightly prefer dynamical DE with $c_{\rm e}\neq 1$ and recently \cite*{Mehrabi:2014ema} and 
\cite{Basilakos:2014yda} have shown that clustering DE reproduces the growth data better in the framework of the 
spherical collapse model. A similar conclusion was suggested also by \cite{Nesseris:2014mfa}.

The growth rate $f=\frac{d\ln \delta_{\rm m}}{d\ln a}$ is usually approximated by $f=\Omega_{\rm m}^{\gamma}$ as 
first introduced by \citet{Peebles1993}. In this parametrization $\gamma$ is the so called growth index and can be 
used to distinguish between DE and modified gravity models 
\citep{Linder:2005in,Huterer:2006mva,Basilakos:2012uu,Rapetti:2012bu}. 
It is well known that for a $\Lambda$CDM model $\gamma$ is independent of redshift and equal to $6/11$. The evolution 
of the matter density $\Omega_{\rm m}$ depends on the evolution of the Hubble parameter $H(a)$ and hence on the 
particular cosmological model adopted. In this paper we consider two distinct models, a constant $w_{\rm d}$ and a 
dynamical $w_{\rm d}(z)$, and we consider $c_{\rm eff}$ as a free parameter. 
Then based on the linear regime we numerically solve the perturbed general relativity (GR) equations to evaluate the 
growth rate of matter in the presence of DE clustering. Using a Markov Chain Monte Carlo (MCMC) method we can 
constrain the cosmological parameters using SnIa, BAO, CMB shift parameter, the Hubble parameter, the Big Bang 
Nucleosynthesis (BBN) and growth rate data $f\sigma_8(z)$. 

The structure of this paper is as follows. In section~\ref{sec1} we derive the equations governing the linear growth 
of matter perturbations in a general relativistic framework and show the effects of DE clustering on the growth rate 
of matter. 
In section~\ref{sec2} we present all the details of the observational data used in this work to constrain the 
cosmological parameters including the DE sound speed and their uncertainties. 
In section~\ref{growth-index}, we provide for the first time (to our knowledge) an approximated solution of the 
growth index of matter fluctuations as a function of the cosmological parameters, DE perturbations and $c_{\rm e}$. 
Finally in section~\ref{sec:con} we conclude and discuss our results.

\section{Effect of dark energy sound speed on the growth rate of matter perturbations}\label{sec1}
In this section we revise the fundamental equations necessary to our analysis. The sound horizon of DE with 
effective sound speed $c_{\rm e}$ in a FRW universe is given by:
\begin{equation}\label{equation:sound-horizon}
 \lambda_{\rm s}(a)=\int_{a_{\rm i}}^{a}\frac{c_{\rm e}(x)}{x\mathcal{H}(x)}~dx\;,
\end{equation}
where $\mathcal{H}=\frac{a^{\prime}}{a}$, the prime being the derivative with respect to conformal time ($\eta$) and 
$a_{\rm i}$ an initial scale factor. The nominal Hubble parameter is given by $H=\frac{\dot{a}}{a}$ and thus 
$\mathcal{H}=aH$ which implies
\begin{equation}\label{HHa}
 \frac{\mathcal{H}^{\prime}}{\mathcal{H}^2}=1+\frac{\dot{H}}{H^{2}}\;,
\end{equation}
where an overdot refers to a derivative with respect to the cosmic time ($t$). In the case of $c_{\rm e}\simeq 1$, 
pressure suppresses any DE perturbation with the consequence that DE may cluster only on scales comparable to the 
horizon.

The opposite situation holds if $c_{\rm e}\ll 1$. Indeed in this case DE can cluster in analogy to the DM component 
and perturbations will grow with time. DE clustering modifies the evolution of DM perturbation and thus it affects 
the rate of structure formation in the universe.

We start our derivation of the relevant equations by considering the line element of an expanding universe in the 
Newtonian gauge without anisotropic stress:
\begin{equation}\label{eq:line-element}
 ds^2=-(1+2\phi)dt^2+a^2(t)(1-2\phi)d\vec{x}^2\;,
\end{equation}
where $\phi$ is the Bardeen potential. 
First-order Einstein equations in Fourier space are:
\begin{eqnarray}
 3\mathcal{H}\phi^{\prime}+\left(3\mathcal{H}^2+k^2\right)\phi & = & 
 -\frac{3\mathcal{H}^2}{2}(\Omega_{\rm m}\delta_{\rm m}+\Omega_{\rm d}\delta_{\rm d})\;,
 \label{eq:first-order-En-eq1}\\
 \phi^{\prime\prime}+3\mathcal{H}\phi^{\prime}+
 \left(\frac{2a^{\prime \prime}}{a}-\mathcal{H}^2\right)\phi & = &
 \frac{3\mathcal{H}^2}{2}\Omega_{\rm d}\frac{\delta p_{\rm d}}{\delta\rho_{\rm d}}\delta_{\rm d}\;,
 \label{eq:first-order-En-eq2}
\end{eqnarray}
where $\Omega_{\rm m}=\Omega_{\rm DM}+\Omega_{\rm b}$ ($\Omega_{\rm d}=1-\Omega_{\rm m}$) is the matter 
(dark energy) density parameter and $\delta_{\rm m}$ ($\delta_{\rm d}$) is the corresponding density contrast. 
The first-order energy-momentum conservation equations for a generic fluid with equation-of-state parameter $w$ are 
\citep{Ma:1995ey}
\begin{eqnarray}
 \delta^{\prime} & = & -(1+w)(\theta-3\phi^{\prime})-
 3\frac{a^{\prime}}{a}\left(\frac{\delta p}{\delta\rho}-w\right)\delta\;,\label{eq:first-order-conser1}\\
 \theta^{\prime} & = & -\frac{a^{\prime}}{a}(1-3w)\theta-\frac{w^{\prime}}{1+w}\theta+
 \frac{\frac{\delta p}{\delta \rho}}{1+w}k^2\delta+k^2\phi\;.\label{eq:first-order-conser2}
\end{eqnarray}
These equations are correct for any fluid with $p=w\rho$ (for dust $w=0$ and for dark energy $w=w_{\rm d}$), where 
$\delta$ is the density contrast, $\theta$ is the divergence of the fluid velocity ($\theta=ik^iv_i$) and 
$\frac{\delta p}{\delta\rho}$ can be written as \citep{Bean:2003fb}
\begin{equation}\label{eq:c_eff}
 \frac{\delta p}{\delta\rho}=c_{\rm e}+3\mathcal{H}(1+w)(c_{\rm e}-c_{\rm ad}^2)
 \frac{\theta}{\delta}\frac{1}{k^2}\;,
\end{equation}
where $c_{\rm a}^2=c_{\rm a}$ is the DE adiabatic sound speed:
\begin{equation}\label{eq:c_a}
 c_{\rm a}=w-\frac{w^{\prime}}{3\mathcal{H}(1+w)}\;.
\end{equation}
Note that the second term in the right hand side of Eq.~(\ref{eq:c_eff}) appears because we demand pressure 
perturbations to be a gauge invariant quantity \citep{Bean:2003fb}. For a perfect fluid, perturbations in the pressure 
are purely determined by the adiabatic sound speed but for an imperfect fluid dissipative processes generate entropic 
perturbations and therefore we have a more general relation. In this case, $c_{\rm e}$ acts like a proxy for pressure 
perturbations and the growth of perturbation in the DE component depends on the effective sound speed and not on the 
adiabatic sound speed any more. In the following this statement will be confirmed by solving the perturbed equations 
numerically.

To study the effect of the DE sound speed on structure formation, we consider a universe with pressure-less DM and a 
DE component with varying equation of state that we specialise to $w_{\rm d}(z)=w_0+w_1\frac{z}{1+z}$. The latter 
parametrization 
is the well known Chevallier-Polarski-Linder (CPL) parametrization \citep{Chevallier:2000qy,Linder:2002et}. 
We eliminate $\theta$ from Eqs.~(\ref{eq:first-order-conser1}) and~(\ref{eq:first-order-conser2}) and find two second 
order differential equations for the density contrast of DM and DE. 
In addition using $\frac{d}{d\eta}=a\mathcal{H}\frac{d}{da}$ and 
$\frac{d^2}{d\eta^2}=a^2\mathcal{H}^2\frac{d^2}{da^2}+(a\mathcal{H}^2+a\dot{\mathcal{H}})\frac{d}{da}$, these 
equations can be written in terms of the scale factor. Finally our desired equations governing the growth of DM and DE 
perturbations are:
\begin{eqnarray}
\frac{d^{2}\delta_{\rm m}}{da^{2}}+ A_{\rm m}\frac{d\delta_{\rm m}}{da}+
B_{\rm m}\delta_{\rm m} & = & S_{\rm m}\;, \label{eq:sec-ord-delta_m}\\
\frac{d^{2}\delta_{\rm d}}{da^{2}}+ A_{\rm d}\frac{d\delta_{\rm d}}{da}+
B_{\rm d}\delta_{\rm d} & = & S_{\rm d}\;, \label{eq:sec-ord-delta_d}
\end{eqnarray}
and the coefficients [see also Eq.~(\ref{HHa})] are:
\begin{eqnarray}\label{eq:cof}
A_{\rm m} & = & \frac{1}{a}\left(2+\frac{\mathcal{H}^{\prime}}{\mathcal{H}^2}\right)
=\frac{1}{a}\left(3+\frac{\dot{H}}{H^2}\right)\;,\\ \nonumber
B_{\rm m} & = & 0\;, \\ \nonumber
S_{\rm m} & = & 3\frac{d^{2}\phi}{da^{2}}+\frac{3}{a}\left[2+\frac{\mathcal{H}^{\prime}}{\mathcal{H}^2}\right]
\frac{d\phi}{da}-\frac{k^2}{a^2\mathcal{H}^2}\phi\;,\\ \nonumber
A_{\rm d} & = & \frac{1}{a}\left[2+
                \frac{\mathcal{H}^{\prime}}{\mathcal{H}^2} +3c_{\rm a}-6w_{\rm d}\right]\;,\\ \nonumber
B_{\rm d} & = & \frac{1}{a^2}\left[3\left(c_{\rm e}-w_{\rm d}\right)
                \left(1+\frac{\mathcal{H}^{\prime}}{\mathcal{H}^2}-
                3w_{\rm d}+3c_{\rm a}-3c_{\rm e}\right)\right.\; \\  \nonumber      
          & + & \left.   \frac{k^2 }{\mathcal{H}^2}c_{\rm e}-3a\frac{dw_{\rm d}}{da}\right]\;,\\ \nonumber
S_{\rm d} & = & (1+w_{\rm d})\left[3\frac{d^{2}\phi}{da^{2}}+\frac{3}{a}\left(2+
\frac{\mathcal{H}^{\prime}}{\mathcal{H}^2}-3c_{\rm a}\right)\frac{d\phi}{da}\right.\; \\  \nonumber 
          & - & \left.\frac{k^2}{a^2\mathcal{H}^2}\phi +\frac{3}{1+w_{\rm d}}\frac{d\phi}{da}
                \frac{dw_{\rm d}}{da}\right]\;,\\ \nonumber
\end{eqnarray}
where $\frac{\mathcal{H}^{\prime}}{\mathcal{H}^2}$ (or $\frac{\dot H}{H^{2}}$) is a function of the scale factor and 
using Friedmann equations we have
\begin{equation}\label{hdot-over-h2}
 \frac{\mathcal{H}^{\prime}}{\mathcal{H}^2}=
 -\frac{1}{2}\frac{\Omega_{\rm m}+\Omega_{\rm d}(1+3w_{\rm d})}{\Omega_{\rm m}+\Omega_{\rm d}}
 =-\frac{1}{2}(1+3\Omega_{\rm d}w_{\rm d})\;,
\end{equation}

These equations are not in agreement with Eq. (44) in \cite{Abramo:2008ip}, which were obtained in the limit of a 
matter dominated universe ($\frac{\mathcal{H}^{\prime}}{\mathcal{H}^2}=-\frac{1}{2}$) and a constant $w_{\rm d}$. 
To resolve this discrepancy, see appendix~(\ref{app:dp}).

We integrate Eqs.~(\ref{eq:sec-ord-delta_m}) and~(\ref{eq:sec-ord-delta_d}) numerically from $z_{\rm i}=100$ to $z=0$, 
in order to obtain the density contrast of DM and DE. 
We use the same procedure of \cite{Abramo:2008ip} to find the initial conditions. In the matter dominated era 
$\phi^{\prime}\simeq 0$, so from Eq.~(\ref{eq:first-order-En-eq1}) we have:
\begin{equation}\label{ini:delm}
 \delta_{\rm m,i}=-2\phi_{\rm i}\left(1+\frac{k^2}{3\mathcal{H_{\rm i}}^2}\right)\;, 
\end{equation}
for the initial value of $\delta_{\rm m}$ and 
\begin{equation}\label{ini:delmp}
 \frac{d\delta_{\rm m,i}}{da}=-\frac{2}{3}\frac{k^2}{\mathcal{H_{\rm i}}^2}\phi_{\rm i}\;,
\end{equation}
for its derivative. 
For $\delta_{\rm d}$ the initial value is set using the adiabatic perturbations condition 
\citep{Kodama:1984,Amendola:2010}, 
\begin{equation}\label{ini:deld}
 \delta_{\rm d,i}=(1+w_{\rm d})\delta_{\rm m,i}\;,
\end{equation}
and its derivative is set to
\begin{equation}\label{ini:deldp}
 \frac{d\delta_{\rm d,i}}{da}=(1+w_{\rm d})\frac{d\delta_{\rm m,i}}{da}+\frac{dw_{\rm d}}{da}\delta_{\rm m,i}\;. 
\end{equation}
According to the above argument, by fixing the initial condition of $\phi_{\rm i}$ we have all the initial conditions. 
We set $\phi_{\rm i}=-6\times10^{-7}$ which corresponds to $\delta_{\rm m}=0.1$ at present time for $k=0.1hMpc^{-1}$. 
Our results are robust under small changes of the initial conditions, and we don't worry about the exact values.
(For  $\phi_{\rm i}=-7\times10^{-8}$, $\delta_{\rm m}$ reach to $0.01$ at present time but $f\sigma_8$ differs less 
than $10^{-4}\%$.)

DE clustering affects the growth of matter perturbations through the change of the potential $\phi$. 
As we noticed the amount of DE clustering is directly related to its effective sound speed. 
We restrict our analysis to the choice of $k=1/\lambda=0.1h$Mpc$^{-1}$ which corresponds to $\lambda=10h^{-1}$Mpc 
\citep{Zhang:2012}. Note that the power-spectrum normalization $\sigma_{8}$ which is the rms mass fluctuation on a 
scale $R_{8}= 8h^{-1}$Mpc corresponds to $k=0.125h$Mpc$^{-1}$. On the other hand it has been common practice to 
assume that the shape of the power spectrum recovered from galaxy surveys matches the linear matter power spectrum 
shape on scales $k\le 0.15h$Mpc$^{-1}$ \citep{Smith:2003,Tegmark:2003ud,Percival:2007}. 
Obviously the choice of $k=0.1h$Mpc$^{-1}$ assures that we are in the linear regime. We find that small variations 
around this value do not really affect the qualitative evolution of the growth rate of clustering and thus of 
$\gamma(z)$.\footnote{Since we are in the linear regime we verify that for different values of $k$ the 
differences in $f\sigma_{8}$ are practically negligible ($\sim 10^{-5}\%$).}

To compare these results with observations we calculate the growth factor 
$f(z)=-\frac{1+z}{\delta_m(z)}\frac{d\delta_m(z)}{dz}$ and the growth index 
$\gamma(z)=\frac{d\ln f(z)}{d\Omega_{\rm m}(z)}$ using our numerical results. The growth index in the $\Lambda$CDM 
model is redshift-independent and approximately equal to $\gamma=0.55$. 
To compare this model to observational data we need to evaluate $f(z)\sigma_8(z)$ where $\sigma_8(z)$ is the mass 
variance in a sphere of radius of 8 Mpc/h. The variance $\sigma_8(z)$ can be written in terms of $\sigma_8$ at 
present time as $\sigma_8(z)=\sigma_{8}(z=0)\frac{\delta_m(z)}{\delta_m(z=0)}$. Also, in order to treat 
$\sigma_{8}\equiv \sigma_{8}(z=0)$ properly for the DE models we rescale the value of $\sigma_{8}$ by 
$\sigma_{8}=\frac{\delta_{m}(z=0)}{\delta_{m,\Lambda}(z=0)}\sigma_{8,\Lambda}$. 
Regarding $\sigma_{8,\Lambda}$ we utilise $\sigma_{8,\Lambda}=0.818\left(0.30/\Omega_{\rm m}\right)^{0.26}$ 
provided by the {\em Planck} analysis of \cite{Spergel:2013rxa} and it  
is also in agreement with the results of {\em Planck 2015} \citep{Planck:2015xua}.

\begin{figure}
\centering
 \includegraphics[width=0.5\textwidth]{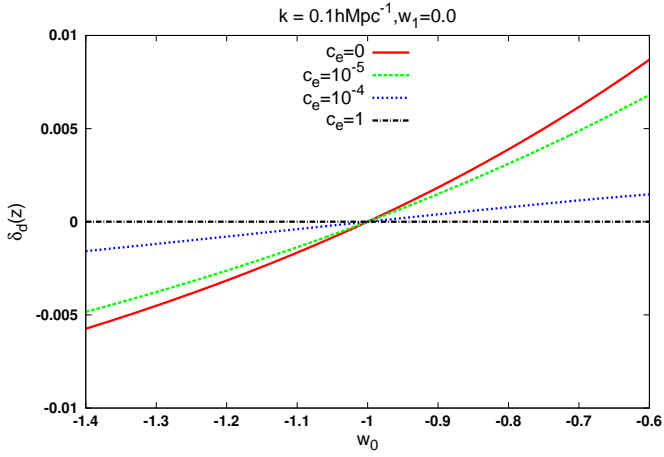}
 \caption{The density contrast of DE as a function of the EoS at the present time for four different values of the 
 sound speed. The red solid curve shows a fully clustering DE model with $c_{\rm e}=0$. The green dashed 
 (blue dotted) curve is for $c_{\rm e}=10^{-5}$ ($c_{\rm e}=10^{-4}$). A non-clustering model with $c_{e}=1$ is shown 
 by a black dashed-dotted line.}
 \label{fig:deld}
\end{figure}

\begin{figure}
\centering
 \includegraphics[width=0.5\textwidth]{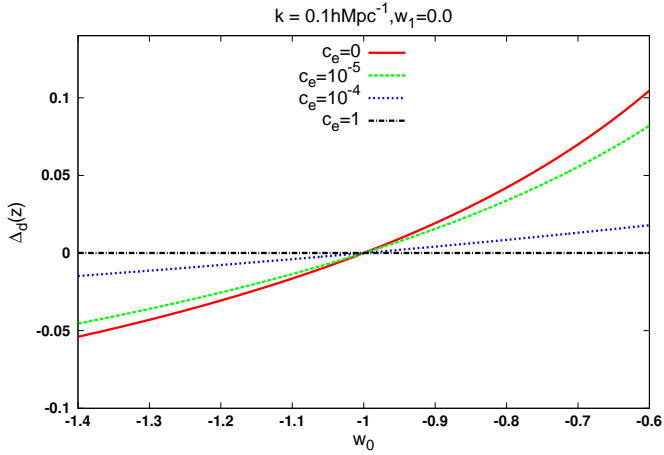}
 \caption{The relative density contrast of DE as a function of the EoS at the present time for four different values of the 
 sound speed. Line style and colours are as in Fig.~(\ref{fig:deld}).}
 \label{fig:Delta}
\end{figure}

DE perturbations not only depend on the sound speed but also on the EoS $w_{\rm d}$. In the limit 
$w_{\rm d}\rightarrow-1$ all DE perturbations are washed out due to the $1+w_{\rm d}$ factor in front of the source 
term in the evolution equation of $\delta_{\rm d}$. To show how the DE sound speed affects the linear evolution of 
DM, we consider $\Omega_{m}=0.28$ and $h=0.7$ in the wCDM model to evaluate $\delta_{\rm d}$ and 
$\Delta_{\rm d}=\frac{\delta_{\rm d}}{\delta_{\rm m}}$, the relative DE density contrast, for a few distinct values of 
the DE sound speed as a function of the EoS. In Fig.~(\ref{fig:deld}) the density contrast of DE as a function of 
$w_{\rm d}$ at the present time is presented. The non-clustering case remains homogeneous but for small values of the 
DE sound speed, the density contrast grows while increasing the EoS. In contrast to the non-clustering case, the fully 
clustering regime with $c_{\rm e}=0$ gives a maximum value for the DE density contrast. 
In Fig.~(\ref{fig:Delta}) the relative DE density contrast is shown as a function of EoS. The behaviour of this 
quantity is similar to that of the density contrast.

\begin{figure}
\centering
 \includegraphics[width=0.5\textwidth]{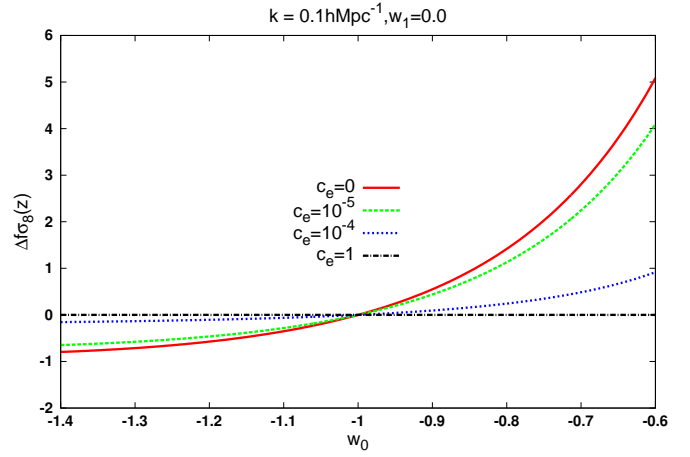}
 \caption{The relative difference of $f\sigma_{8}$ at the present time as a function of EoS. 
 Line style and colours are as in Fig.~(\ref{fig:deld}).}
 \label{fig:delta-fs8}
\end{figure}

\begin{figure}
\centering
 \includegraphics[width=0.5\textwidth]{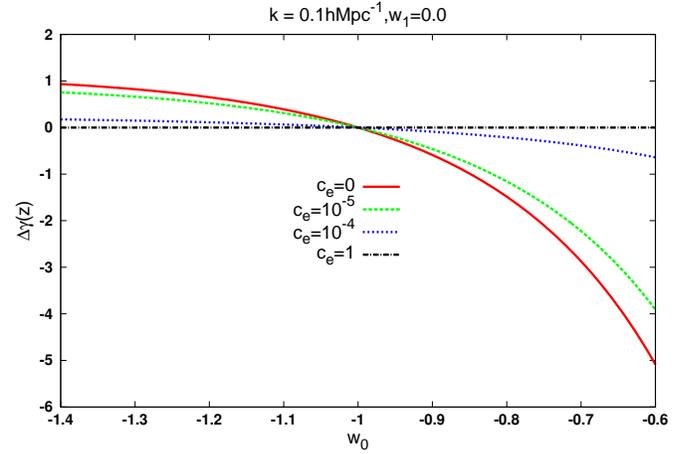}
 \caption{The relative difference of the growth index at present time as a function of EoS. 
 Line style and colours are as in Fig.~(\ref{fig:deld}).}
 \label{fig:grow-index}
\end{figure}

As we stated the quantity $f\sigma_{8}(z)$ is affected by DE clustering. To show how $f\sigma_{8}(z)$ changes with 
the DE sound speed, we evaluate 
$\Delta f\sigma_{8}(z)=\frac{f_{h}\sigma_{8,h}(z)-f\sigma_{8}(z)}{f\sigma_{8}(z)}\times 100$ and 
$\Delta\gamma(z)=\frac{\gamma_{h}(z)-\gamma(z)}{\gamma(z)}\times 100$ as a function of the EoS parameter. In the 
previous equations, $h$ stands for homogeneous DE. For the growth rate, results at present time are presented in 
Fig.~(\ref{fig:delta-fs8}). As expected, the deviation increases by increasing the EoS and for $w_{\rm d}<-0.9$ the 
difference is less than $1\%$. The relative difference between homogeneous and clustering DE for the growth index 
$\Delta\gamma(z=0)$ has been shown in Fig.~(\ref{fig:grow-index}). The difference between the homogeneous and the 
clustering DE models is also very small for $w_{\rm d}$ very close to the $\Lambda$CDM model.

\section{Observational constraints on the dark energy sound speed}\label{sec2}
In this section we use current available observational data sets to constrain the cosmological background parameters 
and the DE sound speed. In this analysis we assume that the DE sound speed is constant in time, regardless of the 
particular equation-of-state parameter adopted. Our cosmological model will be described by the following parameters: 
$\Omega_{\rm m0}$ (matter density), $\Omega_{\rm b0}$ (baryon density), $h=H_{0}/100$ (normalised Hubble constant), 
$w_0$ and $w_1$ (dark energy equation-of-state parameters) and $c_{\rm e}$ (effective sound speed) to describe 
the dark energy perturbations. 
In our analysis we assume a flat universe so that $\Omega_{\rm DM}+\Omega_{\rm b}+\Omega_{\rm d}=1$, hence the amount 
of dark energy is known from the knowledge of the matter and baryon density parameters.

The first data set we consider is the SnIa distance module from Union 2.1 sample \citep{Union2.1:2012}. This data set 
includes 580 SnIa and its $\chi^2$ is given by:
\begin{equation}\label{eq:xi2-sn}
 \chi^{2}_{\rm sn}=\sum_i\frac{[\mu_{\rm th}(z_i)-\mu_{\rm ob}(z_i)]^2}{\sigma_i^2}\;,
\end{equation}
where $\mu_{\rm th}(z)=5\log_{10}\left[(1+z)\int_0^z\frac{dx}{E(x)}\right]+\mu_0$, $\mu_0=42.384-5\log_{10}h$ and 
$\sigma_i$ are the corresponding uncertainties. Before finding the minimum of $\chi^2_{\rm sn}$ we can expand 
$\chi^2_{\rm sn}$ around $\mu_0$
\begin{equation}\label{eq:expand-xi2}
 \chi^2_{\rm sn}=A+2B\mu_0+C\mu_0^2\;,
\end{equation}
where
\begin{eqnarray}\label{eq:xi2-A-B-C}
 A &=& \sum_i\frac{[\mu_{\rm th}(\mu_0=0)-\mu_{\rm ob}]^2}{\sigma_i^2}\;,\nonumber \\
 B &=&\sum_i\frac{[\mu_{\rm th}(\mu_0=0)-\mu_{\rm ob}]}{\sigma_i^2}\;,\nonumber \\
 C &=& \sum_i\frac{1}{\sigma_i^2}\;. \nonumber
\end{eqnarray}
Obviously, for $\mu_{0}=-B/C$ Eq.~(\ref{eq:expand-xi2}) has a minimum, namely $A-\frac{B^2}{C}$. 
Now by defining $\tilde{\chi}_{\rm sn}^2=A-\frac{B^2}{C}$, we can use the minimum of $\tilde{\chi}_{\rm sn}^2$ 
which is independent of $\mu_{0}$ in order to find the best values of the parameters. 
Of course both estimators provide the same results \citep{Nesseris:2005ur}.

The second data set we consider is the BAO sample which includes 6 distinct measurements of the baryon acoustic scale. 
These 6 data points and their references are summarised in Tab.~(\ref{tab:bao}).

\begin{table}
 \centering
 \caption{The current available BAO data which we use in our analysis.}
 \begin{tabular}{c|c|c}
 \hline
 \hline
 $z$     & $d_i$    & Survey \& References  \\ \hline
 $0.106$ & $0.336$  & 6dF \citep{Beutler:2011hx}\\ \hline
 $0.35$   & $0.113$ & SDSS-DR7 \citep{Padmanabhan:2012hf}\\ 
 $0.57$  & $0.073$ & SDSS-DR9 \citep{Anderson:2012sa} \\ \hline
 $0.44$  & $0.0916$ & WiggleZ \citep{Blake:2011en} \\ 
 $0.6$   & $0.0726$ & WiggleZ \citep{Blake:2011en} \\ 
 $0.73$  & $0.0592$ & WiggleZ \citep{Blake:2011en} \\ \hline \hline
 \label{tab:bao}
 \end{tabular}
\end{table}
To find the $\chi^2_{\rm BAO}$ we follow the same procedure as \cite{Hinshaw:2012aka}. So the $\chi^2_{\rm BAO}$ is 
given by
\begin{equation}\label{eq:xi2-bao}
 \chi^2_{\rm BAO}=\mathbf{Y}^{T}\mathbf{C}_{\rm BAO}^{-1}\mathbf{Y}\;,
\end{equation}
where $\mathbf{Y}=(d(0.1)-d_{1},\frac{1}{d(0.35)}-\frac{1}{d_2},\frac{1}{d(0.57)}-
\frac{1}{d_3},d(0.44)-d_{4},d(0.6)-d_{5},d(0.73)-d_{6})$ and
\begin{equation}\label{eq:d(z)}
 d(z)=\frac{r_{\rm s}(z_{\rm drag})}{D_V(z)}\;,
\end{equation}
with
\begin{equation}\label{eq:com-sound-H}
 r_{\rm s}(a)=\int_0^{a}\frac{c_{\rm s}da}{a^2H(a)}\;,
\end{equation}
is the comoving sound horizon at the baryon drag epoch, $c_{\rm s}$ the baryon sound speed and $D_V(z)$ is defined by:
\begin{equation}\label{eq:dv-bao}
 D_V(z)=\left[(1+z)^{2}D^{2}_{\rm A}(z)\frac{z}{H(z)}\right]^{\frac{1}{3}}\;,
\end{equation}
and $D_{\rm A}(z)$ is the angular diameter distance. 
We used the fitting formula for $z_{\rm d}$ from \cite{Eisenstein:1997ik} and the baryon sound speed is given by:
\begin{equation}\label{eq:bary-soun}
 c_{\rm s}(a)=\frac{1}{\sqrt{3(1+\frac{3\Omega_b^0}{4\Omega_{\gamma}^0}a)}}\;,
\end{equation}
where we set $\Omega_{\gamma}^0=2.469\times 10^{-5} h^{-2}$ \citep{Hinshaw:2012aka}. 
The covariance matrix $\mathbf{C}_{\rm BAO}^{-1}$ in Eq.~(\ref{eq:xi2-bao}) was obtained by \cite{Hinshaw:2012aka}
\begin{eqnarray}\label{eq:cij-bao}
 \left(  
\begin{array}{cccccc}
 4444.4 & 0. & 0. & 0. & 0. & 0. \\
 0. & 34.602 & 0. & 0. & 0. & 0. \\
 0. & 0. & 20.6611 & 0. & 0. & 0. \\
 0. & 0. & 0. & 24532.1 & -25137.7 & 12099.1 \\
 0. & 0. & 0. & -25137.7 & 134598.4 & -64783.9 \\
 0. & 0. & 0. & 12099.1 & -64783.9  & 128837.6
\end{array}
\right)\;.\nonumber
\end{eqnarray}

The position of the CMB acoustic peak provides a useful data to constrain dark energy models. 
The position of this peak is given by $(l_{\rm a},R,z_{\ast})$, where $R$ is the scale distance to recombination
\begin{eqnarray}\label{eq:R-cmb}
 l_a & = & \pi\frac{D_{\rm A}(z_{\ast})}{r_s(z_{\ast})}\;,\\
 R & = & \sqrt{\Omega_{\rm m}^{0}}H_{\rm 0}D_{\rm A}(z_{\ast})\;,
\end{eqnarray} 
and $r_{\rm s}(z)$ is the comoving sound horizon defined in Eq.~(\ref{eq:com-sound-H}). In this case we used the 
formula for $z_{\ast}$ from \cite{Hu:1995en}. For the WMAP data set we have \citep{Hinshaw:2012aka}
\begin{equation}\label{eq:x-cmb}
 \mathbf{X}_{\rm CMB}= \left(
\begin{array}{c}
 l_{\rm a}-302.40 \\
 R-1.7264 \\
 z_{\ast}-1090.88
\end{array}
\right)\;,
\end{equation}
and
\begin{eqnarray}
\mathbf{C}_{\rm CMB}^{-1}=\left(
\begin{array}{ccc}
3.182 & 18.253  & -1.429  \\
18.253& 11887.879& -193.808\\
-1.429& -193.808& 4.556
\end{array}
\right)\;.
\end{eqnarray}
In addition to this data set the {\em Planck} data provide more accurate CMB data for which the position of the 
acoustic peak is given by \citep{Shafer:2013pxa}
\begin{equation}\label{eq:x-cmb-pl}
 \mathbf{X}_{\rm CMB}= \left(
\begin{array}{c}
 l_{\rm a}-301.65 \\
 R-1.7499 \\
 z_{\ast}-1090.41
\end{array}
\right)\;,
\end{equation}
and
\begin{eqnarray}
\mathbf{C}_{\rm CMB}^{-1}=\left(
\begin{array}{ccc}
42.7044 & -418.36  & -0.7820  \\
-418.36& 57366.3& -762.152\\
-0.7820& -762.152& 14.6995
\end{array}
\right)\;.
\end{eqnarray}
In both cases 
the $\chi^2_{\rm CMB}$ is given by :
\begin{equation}\label{eq:xi2-cmb}
 \chi^2_{\rm CMB}=\mathbf{X}_{\rm CMB}^{T}\mathbf{C}_{\rm CMB}^{-1}\mathbf{X}_{\rm CMB}\;.
\end{equation}

A further data set used in this work is the Hubble evolution data obtained from the evolution of galaxies 
\citep{Simon:2004tf}. We use the 12 available data points and the $\chi^2$ for this data set is:
\begin{equation}\label{eq:xi2-H}
 \chi^2_{\rm H}=\sum_i\frac{[H(z_i)-H_{\rm ob,i}]^2}{\sigma_i^2}\;.
\end{equation}
The Big Bang Nucleosynthesis (BBN) provides a data point \citep{Serra:2009yp,Burles:2000zk} which constrains mostly 
$\Omega_{\rm b}^{0}$. The $\chi^2_{\rm BBN}$ is given by
\begin{equation}\label{eq:xi2-bbn}
 \chi^2_{\rm BBN}=\frac{(\Omega_{\rm b}^{0}h^2-0.022)^2}{0.002^2}\;.
\end{equation}
The final data set used is the growth rate data. These data were derived from redshift space distortions from galaxy 
surveys including PSCs, 2DF, VVDS, SDSS, 6dF, 2MASS, BOSS and WiggleZ and the data with their references are shown in 
Tab.~\ref{tab:fsigma8data}. We solve Eqs.~(\ref{eq:sec-ord-delta_m}) and~(\ref{eq:sec-ord-delta_d}) numerically to 
find $f(z)\sigma_8(z)$ and compute $\chi^2_{\rm fs}$ with
\begin{equation}\label{eq:xi2-fs}
 \chi^2_{\rm fs}=\sum_i\frac{[f\sigma_8(z_i)-f\sigma_{8,\rm ob}]^2}{\sigma_i^2}\;.
\end{equation}

\begin{center}
\begin{table}
\caption{The $f\sigma_8(z)$ data points including their references and surveys.}
\begin{tabular}{ccc}
\hline
\hline
z & $f\sigma_8(z)$ & Reference \\
\hline
$0.02$  & $0.360\pm0.040$ & \cite{Hudson:2012gt}\\
$0.067$ & $0.423\pm0.055$ & \cite{Beutler:2012px}\\
$0.10$  & $0.37\pm0.13$   & \cite{Feix:2015dla}\\
$0.17$  & $0.510\pm0.060$ & \cite{Percival:2004fs}\\
$0.35$  & $0.440\pm0.050$ & \cite{Song:2008qt,Tegmark:2006az}\\
$0.77$  & $0.490\pm0.180$ & \cite{Guzzo:2008ac,Song:2008qt}\\
$0.25$  & $0.351\pm0.058$ & \cite{Samushia:2011cs}\\
$0.37$  & $0.460\pm0.038$ & \cite{Samushia:2011cs}\\
$0.22$  & $0.420\pm0.070$ & \cite{Blake:2011rj}\\
$0.41$  & $0.450\pm0.040$ & \cite{Blake:2011rj}\\
$0.60$  & $0.430\pm0.040$ & \cite{Blake:2011rj}\\
$0.60$  & $0.433\pm0.067$ & \cite{Tojeiro:2012rp}\\
$0.78$  & $0.380\pm0.040$ & \cite{Blake:2011rj}\\
$0.57$  & $0.427\pm0.066$ & \cite{Reid:2012sw}\\
$0.30$  & $0.407\pm0.055$ & \cite{Tojeiro:2012rp}\\
$0.40$  & $0.419\pm0.041$ & \cite{Tojeiro:2012rp}\\
$0.50$  & $0.427\pm0.043$ & \cite{Tojeiro:2012rp}\\
$0.80$  & $0.47\pm0.08$   & \cite{delaTorre:2013rpa}\\
\hline
\hline
\end{tabular}
\label{tab:fsigma8data}
\end{table}
\end{center}

The overall likelihood function is given by the product of the individual likelihoods:
\begin{equation}\label{eq:like-tot}
 {\cal L}_{\rm tot}={\cal L}_{\rm sn} \times {\cal L}_{\rm BAO} \times {\cal L}_{\rm CMB} \times {\cal L}_{H} \times 
 {\cal L}_{\rm \rm BBN} \times {\cal L}_{\rm fs}\;,
\end{equation}
and the total chi-square $\chi^2_{\rm tot}$ is given by:
\begin{equation}\label{eq:like-tot_chi}
 \chi^2_{\rm tot}=\chi^2_{\rm sn}+\chi^2_{\rm BAO}+\chi^2_{\rm CMB}+\chi^2_{H}+\chi^2_{\rm BBN}+\chi^2_{\rm fs}\;.
\end{equation}

\begin{table}
 \begin{center}
  \caption{The best value parameters and their 1-$\sigma$ uncertainty for the wCDM model.}
  \begin{tabular}{| c | c |c |}
   \hline
   \hline
   Parameters          & Best (WMAP) & Best (Planck) \\
   \hline
   $h$                 & $0.6955^{+0.0040}_{-0.0037}$ &$0.7064^{+0.0011}_{-0.0012}$\\
   \hline
   $\Omega_{\rm DM}^0$ & $0.2273^{+0.0027}_{-0.0029}$ &$0.2361^{+0.0010}_{-0.0010}$\\
   \hline
   $\Omega_{\rm b}^0$  & $0.0470^{+0.0004}_{-0.0005}$ &$0.0482^{+0.0003}_{-0.0002}$\\
   \hline
   $w_0$               & $-0.9436^{+0.0144}_{-0.0141}$ &$-0.9975^{+0.0055}_{-0.0053}$\\
   \hline
   $c_{\rm e}$         & $0.$ & $0.001$\\
   \hline
    $\sigma_{\rm 8}$         & $0.837$ & $0.829$\\
   \hline
   \hline
   \hline
  \end{tabular}
 \label{tab:res1}
 \end{center}
\end{table}

\begin{table}
 \begin{center}
  \caption{The best value parameters and their 1-$\sigma$ uncertainty for the w(t)CDM model.}
  \begin{tabular}{| c | c |c |}
   \hline
   \hline
    Parameters          & Best (WMAP) & Best (Planck) \\
   \hline
   $h$                & $0.7001^{+0.0040}_{-0.0038}$ & $0.7070^{+0.0012}_{-0.0013}$\\
   \hline
   $\Omega_{\rm DM}^0$ & $0.2234^{+0.0028}_{-0.0027}$ & $0.2361^{+0.0012}_{-0.0011}$\\
   \hline
   $\Omega_{\rm b}^0$ & $0.0474^{+0.0005}_{-0.0005}$ & $0.0481^{+0.0003}_{-0.0003}$ \\
   \hline
   $w_0$              & $-1.0176^{+0.0128}_{-0.0124}$ & $-0.95204^{+0.0060}_{-0.0058}$ \\
   \hline
   $w_1$              & $0.3289^{+0.0395}_{-0.0405}$ & $ -0.18512^{+0.0205}_{-0.0195}$ \\
   \hline
   $c_{\rm e}$        & $0.002$ & $0.$\\
   \hline
    $\sigma_{\rm 8}$        & $0.840$ & $0.829$\\
   \hline
   \hline
   \hline
  \end{tabular}
 \label{tab:res2}
 \end{center}
\end{table}
We calculate the total chi-square $\chi^2_{\rm tot}$ and find the best value of the parameters with an MCMC algorithm. 
The number of degrees of freedom is $\nu=N-n_{\rm fit}-1$, where $N=616$ and $n_{\rm fit}$ is the number of the 
fitted parameters. The results of this analysis for the wCDM, w(t)CDM and $\Lambda$CDM are summarized in 
Tabs.~(\ref{tab:res1}),~(\ref{tab:res2}) and~(\ref{tab:lcdm}) respectively.

\begin{table}
 \begin{center}
  \caption{The best value parameters and their 1-$\sigma$ uncertainty for the $\Lambda$CDM model.}
  \begin{tabular}{| c | c |c |}
   \hline
   \hline
   Parameters          & Best (WMAP) & Best (Planck) \\
   \hline
   $h$                 & $0.7048^{+0.0042}_{-0.0041}$ & $0.7069^{+0.0011}_{-0.0010}$\\
   \hline
   $\Omega_{\rm DM}^0$ & $0.2261^{+0.0030}_{-0.0029}$ & $0.2359^{+0.0010}_{-0.0011}$\\
   \hline
   $\Omega_{\rm b}^0$  & $0.0456^{+0.0006}_{-0.0005}$ & $0.0481^{+0.0003}_{-0.0003}$ \\
   \hline
    $\sigma_{\rm 8}$        & $0.839$ & $0.829$\\
   \hline
   \hline
   \hline
  \end{tabular}
 \label{tab:lcdm}
 \end{center}
\end{table}

To compare the DE models we have computed the corrected Akaike information criterion (AIC) 
\citep{Akaike:1974,Sugiura:1978} which, in our case, due to $N/n_{\rm fit}>40$, is given by:
\begin{equation}\label{eq:AIC}
 {\rm AIC}=\chi^2_{min}+2n_{\rm fit}\;.
\end{equation}
A smaller value of AIC indicates a better model-data fit. Of course it is well known that small differences in AIC are 
not necessarily significant and therefore, in order to assess the effectiveness of the different models in reproducing 
the data, we need to estimate the model pair difference $\Delta$AIC$={\rm AIC}_{y}-{\rm AIC}_{x}$. 
The higher the value of $|\Delta{\rm AIC}|$, the higher the evidence against the model with a higher value of 
${\rm AIC}$. With a difference $|\Delta$AIC$|\ge 2$ indicating a positive evidence and $|\Delta$AIC$|\ge 6$ indicating 
a strong evidence, while a value $|\Delta$AIC$|\le 2$ indicates consistency among the two models. 
The results of our analysis are the following:
\begin{enumerate}
 \item Using WMAP data:
\begin{itemize}
\item For the wCDM model,  $\chi^2_{\rm min}=586.53$, $n_{\rm fit}=5$, so AIC=596.53
\item For the w(t)CDM model,  $\chi^2_{\rm min}=585.32$, $n_{\rm fit}=6$, so AIC=597.32
 \item For the $\Lambda$CDM model,  $\chi^2_{\rm min}=589.22$, $n_{\rm fit}=3$, so AIC=595.32
\end{itemize} 
\item  Using Planck data:
\begin{itemize}
\item For the wCDM model,  $\chi^2_{\rm min}=595.76$, $n_{\rm fit}=5$, so AIC=605.76
\item For the w(t)CDM model,  $\chi^2_{\rm min}=595.50$, $n_{\rm fit}=6$, so AIC=607.50
 \item For the $\Lambda$CDM model,  $\chi^2_{\rm min}=595.79$, $n_{\rm fit}=3$, so AIC=601.79
\end{itemize} 
\end{enumerate}
Concerning the best value of the dark energy sound speed we find 
that it tends to zero 
but the corresponding error bars remain quite large 
within $1\sigma$. In particular $c_{\rm e}$ lies in the range $\in[0.,1]$.

\begin{figure}
\centering
\includegraphics[width=.5\textwidth]{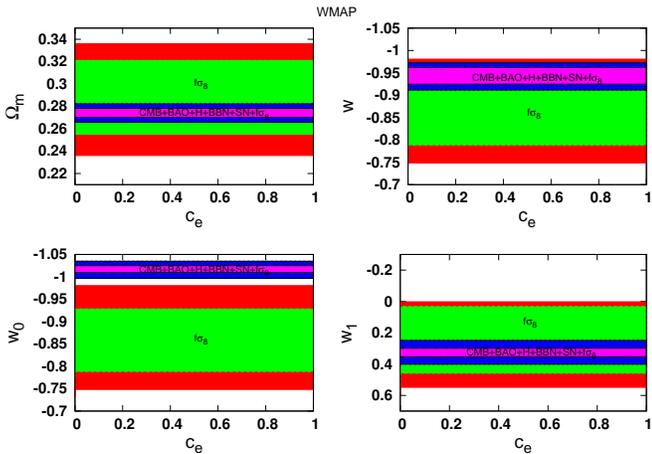}
\caption{The $1\sigma$ and $2\sigma$ contours of $\Omega_{\rm m}$(wCDM), $w$(wCDM), $w_0$(w(t)CDM) and $w_1$(w(t)CDM)  versus DE sound speed using {\em WMAP} data. The $1\sigma$ and $2\sigma$ contours correspond to $\chi^2-\chi^2_{\rm b}=2.3$ and $\chi^2-\chi^2_{\rm b}=6.16$. The green (red) area correspond to $1\sigma$ ($2\sigma$) using only $f\sigma_8$ data and purple (blue) show $1\sigma$ ($2\sigma$) using all data set.}
\label{fig:con-wp}
\end{figure}

\begin{figure}
\centering
\includegraphics[width=.5\textwidth]{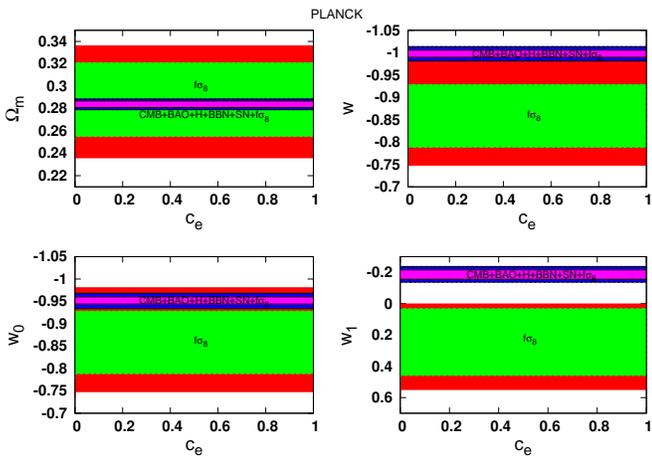}
\caption{Same as Fig.~(\ref{fig:con-wp}) but using the {\em Planck} shift parameter.}
\label{fig:con-pl}
\end{figure}

 In order to investigate the range of validity for 
$c_{\rm e}$, in Figs.~(\ref{fig:con-wp}) and (\ref{fig:con-pl})
we provide the $1\sigma$ and $2\sigma$ contours of our analysis.
Note that in both plots the upper panels are for wCDM in which we present 
the confidence levels in the $(c_{\rm e},\Omega_{\rm m})$ and 
$(c_{\rm e},w)$ planes, where  $\Omega_{\rm m}=\Omega_{\rm DM}^{0}+\Omega_{\rm b}^{0}$.
In the bottom panels of Figs.~(\ref{fig:con-wp}) and (\ref{fig:con-pl}) 
the contours for $w_0$ and $w_{1}$ in the CPL model are 
shown with respect to the DE sound speed.   
From this analysis it becomes clear that there is a strong degeneracy 
between $c_{\rm e}$ and $(\Omega_{\rm m},w)$ which implies that all
values in the interval $0 \le c_{\rm e} \le 1$ are 
acceptable within the $1\sigma$ uncertainty.
As we stated, with data used in this paper the error bar of 
DE sound speed is quite large.  

\begin{figure}
\centering
\includegraphics[width=.5\textwidth]{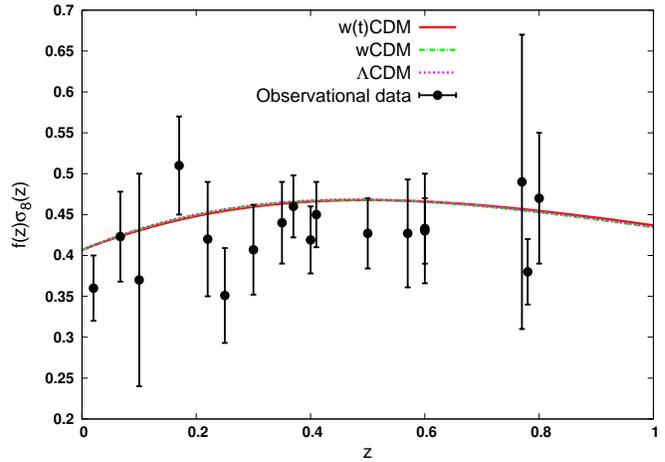}
\caption{The $f\sigma_8(z)$ quantity (using {\em Planck} data), for the best values cosmological
parameters for the wCDM (green dot-dashed curve) and w(t)CDM (red solid curve) models.
The $\Lambda$CDM model is shown by the violet short-dashed curve.}
\label{fig:fs8-exp-pl}
\end{figure}

\begin{figure}
\centering
\includegraphics[width=.5\textwidth]{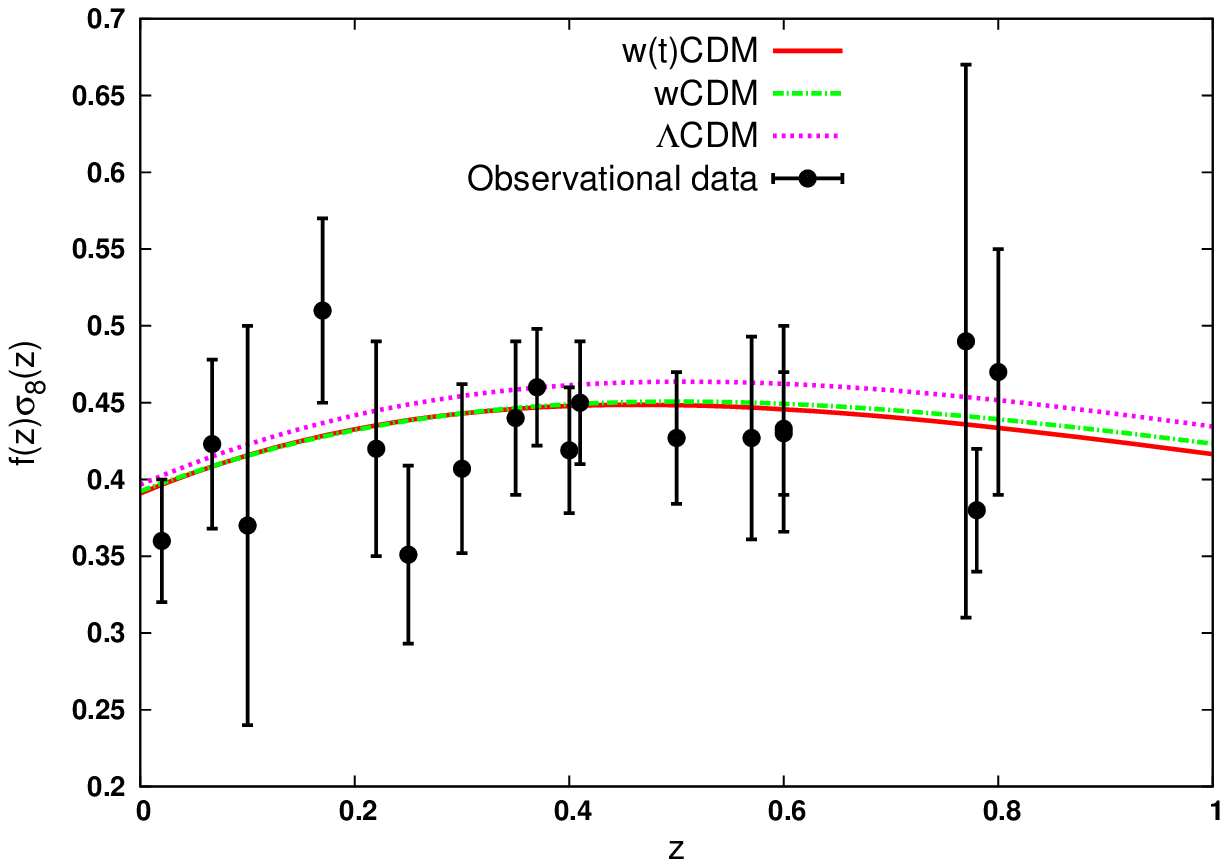}
\caption{The $f\sigma_8(z)$ quantity (using {\em WMAP} data), for the best values cosmological
parameters for the wCDM (green dot-dashed curve) and w(t)CDM (red solid curve) models.
The $\Lambda$CDM model is shown by the violet short-dashed curve.}
\label{fig:fs8-exp-wm}
\end{figure}

\begin{figure}
\centering
\includegraphics[width=.5\textwidth]{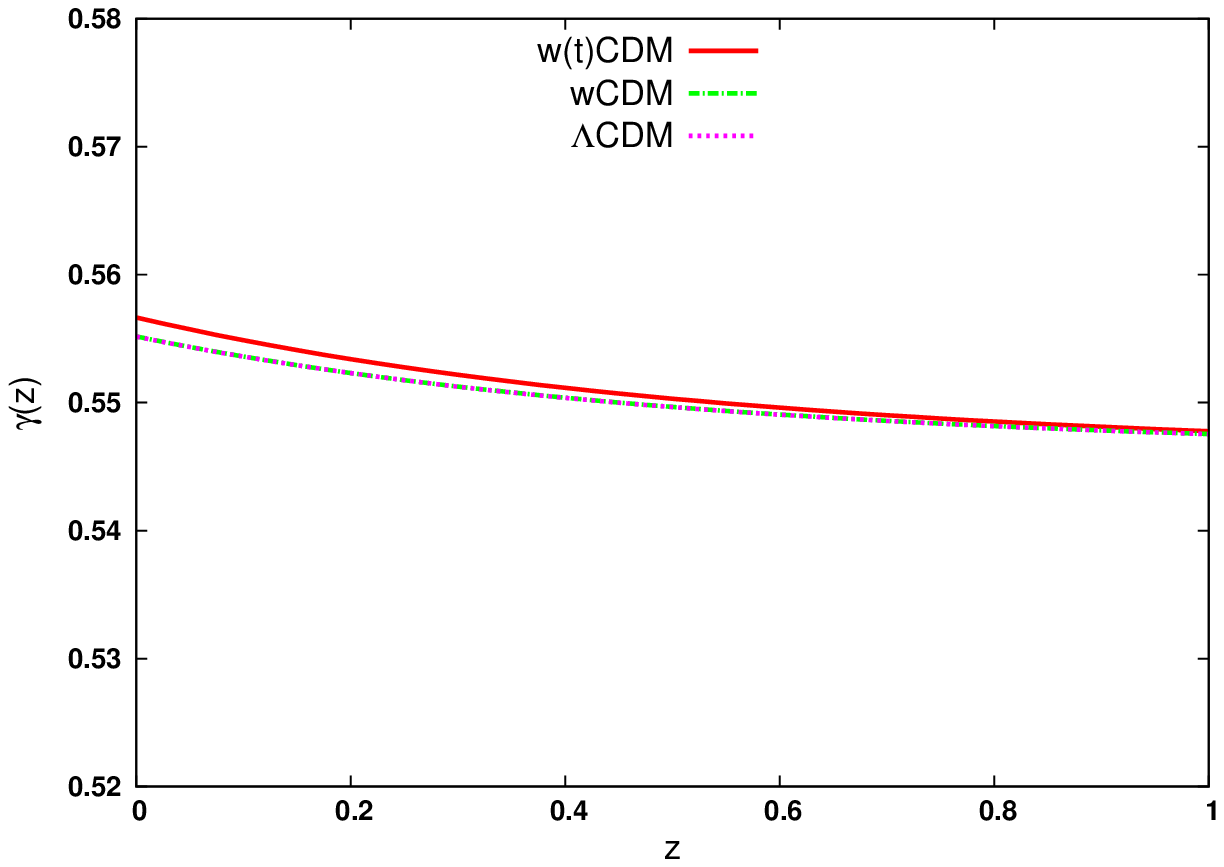}
\caption{The growth index (using {\em Planck} data), for the best values cosmological
parameters for the wCDM (green dot-dashed curve) and w(t)CDM (red solid curve) models.
The $\Lambda$CDM model is shown by the violet short-dashed curve.}
\label{fig:ga-pl}
\end{figure}

\begin{figure}
\centering
\includegraphics[width=.5\textwidth]{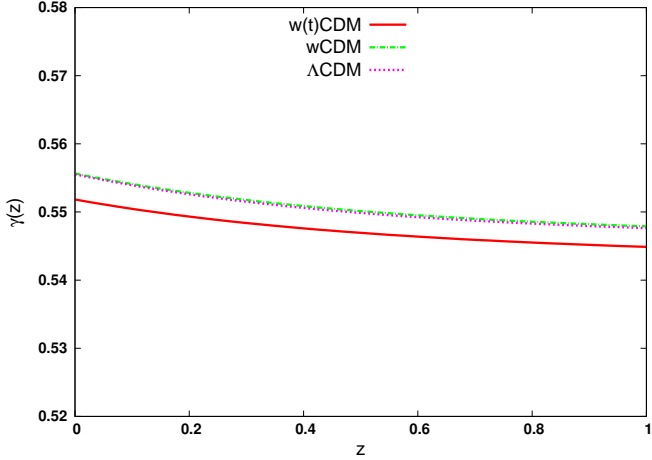}
\caption{The growth index (using {\em WMAP} data), for the best values cosmological
parameters for the wCDM (green dot-dashed curve) and w(t)CDM (red solid curve) models.
The $\Lambda$CDM model is shown by the violet short-dashed curve.}
\label{fig:ga-wm}
\end{figure}

In Figs.~(\ref{fig:fs8-exp-pl}) and (\ref{fig:fs8-exp-wm}) we present the quantity $f\sigma_8(z)$ for our best value 
parameters by considering the {\em Planck} and {\em WMAP} data for the wCDM, w(t)CDM and the $\Lambda$CDM models, 
respectively. We also show the observational data points. In addition to this quantity in Figs.~(\ref{fig:ga-pl}) and 
(\ref{fig:ga-wm}) the growth index for the best values of the parameters have been shown. Note that using 
{\em Planck} CMB data our likelihood analysis indicates that all three models are very close to each others.
\footnote{See the results of $\chi^2$ for the {\em Planck} case.}

 Previous works in literature tried to put constraints on the dark energy effective sound speed $c_{\rm e}$ using different kind of data. \cite{dePutter:2010vy}combination of CMB temperature power spectrum data, their cross-correlation with several mass-density tracers and the SDSS LRG auto-correlation function. Supernovae data were used to break degeneracies with background cosmological parameters. \cite{Hannestad:2005ak}set of Supernova data, LSS and CMB power spectra. Finally \cite{Xia:2007km} performed a similar analysis for a single perfect fluid and a two-field Quintom dark energy model with $w=-1$ crossing by analysing CMB anisotropy data, LSS and SNIa observational data. In all these studies, using a similar approach to the one used in this work, the authors reach our same results. While previous and current data can constrain at a good level the current equation-of-state parameter of the dark energy component, the quality of the observations is unfortunately still not sufficient enough to put any constraint on the dark energy effective sound speed. Note however that this is also due to the negligible contribution of dark energy at early times on one side, and to the fact that current observations favour $w\simeq -1$. As pointed out by \cite{dePutter:2010vy}, if one considers the case of early dark energy models \citep{Doran:2006kp} where the contribution of dark energy at early times, i.e. CB, is not negligible, then more stringent limits can be set on $c_{\rm e}$.

\section{Growth index analytic solution}\label{growth-index}
In section~\ref{sec1} we investigated the evolution of the growth index by solving numerically the system of 
Eqs.~(\ref{eq:first-order-En-eq2}), (\ref{eq:sec-ord-delta_m}) and (\ref{eq:sec-ord-delta_d}). Here our aim is to 
extend the work of \cite{Basilakos:2014yda} in order to provide a general $\gamma(z)$ approximated solution which 
can be used in studies of structure formation. 
On sub-horizon scales, namely $\frac{k^2}{a^2} \gg H^2$ (or $k^{2}\gg \mathcal{H}^2$), Poisson equation (see 
appendix~\ref{app:Poisson}) takes the form
\begin{equation}\label{drhomss}
-\frac{k^2}{a^2}\phi=\frac{3H^{2}}{2}\left[\Omega_{\rm m}\delta_{\rm m}+
\Omega_{\rm d}\delta_{\rm d}(1+3c_{\rm e})\right]\;.
\end{equation}
Under the above conditions, Eq.~(\ref{eq:sec-ord-delta_m}) becomes
\begin{equation}\label{odedelta}
a^{2}\frac{d^{2}\delta_{\rm m}}{da^{2}}+
a\left(3+\frac{\dot{H}}{H^2}\right)\frac{d\delta_{\rm m}}{da}=
\frac{3}{2}\left[\Omega_{\rm m}\delta_{\rm m}+(1+3c_{\rm e})
\Omega_{\rm d}\delta_{\rm d}\right]\;.
\end{equation}
In this framework, for $\delta_{\rm d}=0$, the latter equation reduces to the well known scale independent equation 
which is also valid for the concordance $\Lambda$ cosmology.

Concerning the equation of state parameter, it is well known that one can express it in terms of the Hubble parameter 
\citep{Saini:1999ba,Huterer:2001ez}
\begin{equation}
w_{\rm d}(a)=\frac{-1-\frac{2}{3}a\frac{d\ln H}{da}}{1-\Omega_{\rm m}(a)}\;,
\end{equation}
or 
\begin{equation}\label{eos22}
a\frac{d{\rm ln} H}{da}=\frac{{\dot H}}{H^{2}}=-\frac{3}{2}-\frac{3}{2}w_{\rm d}(a)\Omega_{\rm d}(a)\;,
\end{equation}
where $\Omega_{\rm m}(a)=1-\Omega_{\rm d}(a)=\frac{\Omega_{m0}}{a^{3}E^{2}(a)}$ and
$E(a)=H(a)/H_{0}$. 
Now, substituting Eq.~(\ref{eos22}) and $f=d{\rm ln}\delta_{\rm m}/d{\rm ln}a$ into Eq.~(\ref{odedelta}) we obtain 
the basic differential equation which governs the growth rate of clustering
\begin{equation}\label{fzz222}
 a\frac{df}{da}+f^{2}+\left(\frac{1}{2}-\frac{3}{2}w_{\rm d}\Omega_{\rm d}\right)f
 =\frac{3}{2}\left[\Omega_{\rm m}+ (1+3c_{\rm e})\Delta_{\rm d} \Omega_{\rm d} \right]\;,
\end{equation}
where $\Delta_{\rm d}(a)\equiv \delta_{\rm d}/\delta_{\rm m}$. 
To this end, changing the variables in Eq.~(\ref{fzz222}) from $a(z)$ to redshift 
[$\frac{df}{da}=-(1+z)^{-2}\frac{df}{dz}$] and utilising $f(z)=\Omega_{m}(z)^{\gamma(z)}$ we arrive to
\begin{equation}\label{Poll}
-(1+z)\gamma_{z}{\rm ln}(\Omega_{\rm m})+\Omega_{\rm m}^{\gamma}+
3w_{\rm d}\Omega_{\rm d}\left(\gamma-\frac{1}{2}\right)+\frac{1}{2}=
\frac{3}{2}\Omega_{\rm m}^{1-\gamma}X\;,
\end{equation}
where $\gamma_{z}=d\gamma/dz$ and
\begin{equation}\label{xxx}
X(z)=1+\frac{\Omega_{\rm d}(z)}{\Omega_{\rm m}(z)}\Delta_{\rm d}(z)(1+3c_{\rm e})\;.
\end{equation}

On the other hand, the parametrization 
$f(a)=d{\rm ln}\delta_{\rm m}/d{\rm ln} a \simeq \Omega_{m}(a)^{\gamma(a)}$ 
has a great impact in cosmological studies because it can be used in order to simplify the numerical calculations of 
Eq.~(\ref{odedelta}). Obviously, a direct integration gives
\begin{equation}\label{Dz221}
\delta_{\rm m}(a,\gamma)=a(z) \;{\rm exp} \left[\int_{a_{i}}^{a(z)} \frac{du}{u}\;
\left(\Omega_{\rm m}^{\gamma}(u)-1\right) \right]\;,
\end{equation}
where $a(z)=1/(1+z)$ and $a_{i}$ is the scale factor of the universe at which the matter component dominates the 
cosmic fluid (here we use $a_{i} \simeq 10^{-1}$ or $z_{i}\simeq 10$). 
Hence, the linear growth factor normalised to unity at the present epoch is 
$D(a)=\frac{\delta_{\rm m}(a,\gamma)}{\delta_{\rm m}(1,\gamma)}$. 
Therefore, in order to proceed with the analysis we need to somehow know the functional form of $\gamma(z)$. 
From the phenomenological point of view we may parametrize $\gamma(z)$ as follows
\begin{equation}\label{Param}
\gamma(z)=\gamma_{0}+\gamma_{1}y(z)\;.
\end{equation}
This equation can be seen as a first order Taylor expansion around some cosmological quantity such as $a(z)$ and $z$.

Recently, it has been found \citep[][and references therein]{Basilakos:2012ws,Basilakos:2012uu} that for those $y(z)$ 
functions which satisfy the condition $y(0)=0$ [or $\gamma(0)=\gamma_{0}$], the parameter $\gamma_{1}$ is written 
as a function of $\gamma_{0}$. For example, at the present epoch 
[$z=0$, $\gamma_{z}(0)=\gamma_{1}y_{z}(0)$, $X_{0}=X(0)$, $w_{0}=w_{d}(0)$], Eq.~(\ref{Poll}) is written as  
\begin{equation}\label{Poll2}
\gamma_{1}=\frac{\Omega_{\rm m0}^{\gamma_{0}}+3w_{0}(\gamma_{0}-\frac{1}{2})
\Omega_{\rm d0}+\frac{1}{2}-\frac{3}{2}\Omega_{\rm m0}^{1-\gamma_{0}} X_{0}}
{y_{z}(0)\ln  \Omega_{\rm m0}}\;,
\end{equation}
where $y_{z}=dy/dz$. 
Note that a similar equation has been found in \cite{Basilakos:2014yda} in the case of $c_{\rm e}\equiv w_{\rm d}$ 
with $w_{\rm d}=const$. As it is expected, for the homogeneous DE case ($\Delta_{\rm d}=0$, $X=1$), we verify that 
the above formula boils down to that of \cite{Polarski:2007rr} for $y(z)=z$. Within this framework, assuming 
$y(z)=1-a(z)=\frac{z}{1+z}$ \citep{Ballesteros:2008qk}, we fully recover results in literature 
\citep{Ishak:2009qs,Belloso:2011ms,DiPorto:2012ey}. 
Notice that below we focus on $y(z)=1-a(z)=\frac{z}{1+z}$ with $y_{z}(0)=1$. The fact that 
$\Omega_{\rm d}(z) \simeq 0$ at $z\gg 1$ implies that the asymptotic value of the growth index 
$\gamma_{\infty}=\gamma_{0}+\gamma_{1}$ is not really affected by the dark energy clustering. 
Therefore, plugging $\gamma_{0}=\gamma_{\infty}-\gamma_{1}$ 
\footnote{Regarding the asymptotic value of the growth index we use $\gamma_{\infty}\approx 3(w-1)/(6w-5)$ for the 
wCDM model \citep[see][]{Linder:2007hg,Nesseris:2007pa} and $\gamma_{\infty}\approx 0.55+0.05[1+w(z=1)]$ for the 
w(t)CDM model \citep{Linder:2005in}.} into Eq.~(\ref{Poll2}) we can obtain the constants 
$\gamma_{0,1}$ in terms of $(\Omega_{\rm m0},w_{0},\Delta_{\rm d0},c_{\rm e})$.

\begin{figure}
\centering
 \includegraphics[width=0.5\textwidth]{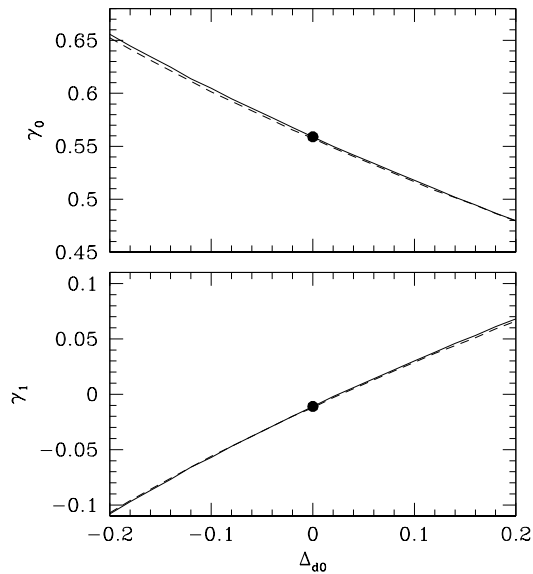}
 \caption{The pair $(\gamma_{0},\gamma_{1})$ as a function of $\Delta_{\rm d0}$. The solid and the dashed lines 
 correspond to the w(t)CDM and wCDM dark energy models, respectively. The homogeneous case $\Delta_{\rm d0}=0$ is 
 shown by the solid point. For the cosmological parameters, we use the values of Table 3 and 4 (third column).}
 \label{fig:delta}
\end{figure}
In Fig.~(\ref{fig:delta}) we present $(\gamma_{0},\gamma_{1})$ as a function of $\Delta_{\rm d0}$. 
The curves are constructed using the parameters from Table 3 and 4 (third column) and they correspond to w(t)CDM 
(solid) and wCDM (dashed) models. We observe that for $\Delta_{\rm d0}>0$ the growth index starts to deviate from 
that of the $\Lambda$CDM model, namely $\gamma_{0}<0.55$ and $\gamma_{1}>0$. In the case of $\Delta_{\rm d0}<0$ the 
value of $\gamma_{0}$ is greater than that of the homogeneous case ($\gamma_{0}>0.55$). In this context, concerning 
the value of $\gamma_{1}$ we find that it becomes negative. Of course for $\Delta_{\rm d0}=0$ the pair 
$(\gamma_{0},\gamma_{1})$ reduces to that of the homogeneous case (see solid points in Fig.~(\ref{fig:delta})), as it 
should.

\section{Conclusions}\label{sec:con}
To summarize, we study the impact of dark energy clustering on the growth index of matter fluctuations. 
Initially we provide the most general form of the equations governing dark matter and dark energy clustering 
within the framework of $c_{\rm e}=const$. Then using the well known equation of state parameters, namely 
$w_{\rm d}(z)=w_{0}+w_{1}z/(1+z)$, $w_{\rm d}(z)=const$ and the current cosmological data we place constrains on the 
cosmological parameters, including that of the 
effective sound speed $c_{\rm e}$. Although the likelihood function 
peaks at $c_{\rm e}\sim 0$, which indicates that the dark energy component  
clusters in analogy to the matter component $c_{\rm e}\sim 0$, the 
corresponding error bars are quite large within 
$1-\sigma$ uncertainties which implies that 
$c_{\rm e}$ remains practically unconstrained. 
We also compared our findings with previous 
work reaching the same conclusion that at the moment 
the quality of cosmological data is not sufficient 
enough to put constraint on the dark energy effective sound speed.
Future cosmological data, based for example on 
{\em Euclid}, are expected to improve even further the relevant 
constraints on $c_{\rm e}$ and thus the validity of 
clustered dark energy will be effectively tested. 
Finally, we have derived a new approximated 
solution of the growth index in terms of the 
cosmological parameters, dark energy perturbations and $c_{\rm e}$. 

\section*{Acknowledgements}
We thank the anonymous referee whose comments helped to improve the paper.

{\footnotesize
\bibliographystyle{mn2e}
\bibliography{ref}
}

\appendix

\section{Proof of Eq. (12)}\label{app:dp}
We start with Eqs.~(\ref{eq:first-order-conser1}) and~(\ref{eq:first-order-conser2}). 
The term $\frac{\delta p}{\delta \rho}$ appears in both equations but it behaves very differently in these 
equations. In the first equation we have
\begin{equation}\label{eq:dp1}
 -3\mathcal{H}\frac{\delta p}{\delta \rho}\delta=-3\mathcal{H}c_e\delta-
 9\frac{\mathcal{H}^2}{k^2}(1+w_{\rm d})(c_e-c_a)\theta\;,
\end{equation}
and on sub-horizon scale we can neglect the latter term ($k^2\gg\mathcal{H}^2$), but in 
Eq.~(\ref{eq:first-order-conser2}) we have
\begin{equation}\label{eq:dp2}
 k^2\frac{\delta p}{\delta \rho}\delta=k^2c_e\delta+3\mathcal{H}(1+w_{\rm d})(c_e-c_a)\theta\;,
\end{equation}
where the latter term can not be neglected. Differentiating Eq.~(\ref{eq:first-order-conser1}) with respect to 
conformal time we have:
\begin{eqnarray}\label{eq:deltapp}
 \delta^{\prime\prime}&+&w_{\rm d}^{\prime}\theta +(1+w_{\rm d})\theta^{\prime}+3 \mathcal{H}^{\prime}c_e\delta\; \\ \nonumber
 &+& 3\mathcal{H}c_e\delta^{\prime}-3\mathcal{H}^{\prime}w_{\rm d}\delta-3 \mathcal{H}w_{\rm d}^{\prime}\delta\;\\ \nonumber
 &-&3 \mathcal{H}w_{\rm d}\delta^{\prime}
 =3w_{\rm d}^{\prime}\phi^{\prime}+ 3(1+w_{\rm d})\phi^{\prime\prime}\;.
\end{eqnarray}
Now from Eq.~(\ref{eq:first-order-conser2})
\begin{equation}\label{eq:thetadot}
 \theta^{\prime}  =  -\mathcal{H}(1-3w_{\rm d})\theta-\frac{w_{\rm d}^{\prime}}{1+w_{\rm d}}\theta+k^2\frac{c_{\rm e}
 \delta}{1+w_{\rm d}} +  3\mathcal{H}(c_{\rm e}-c_{\rm a})\theta+k^2\phi\;,
\end{equation}
and from Eq.~(\ref{eq:first-order-conser1})
\begin{equation}\label{eq:theta}
 \theta = 3\phi^{\prime}-\frac{\delta^{\prime}}{1+w_{\rm d}}-\frac{3\mathcal{H}c_{\rm e}\delta}{1+w_{\rm d}}+
\frac{3\mathcal{H}w_{\rm d}\delta}{1+w_{\rm d}}\;.
\end{equation}
Substituting Eqs.~(\ref{eq:thetadot}) and (\ref{eq:theta}) into Eq.~(\ref{eq:deltapp}),
we have a second order equation governing the evolution of DE. Changing the independent variable to the 
scale factor, the coefficients in Eqs.~(\ref{eq:cof}) can be retrieved. 
On the other hand if we consider $\frac{\delta p}{\delta \rho}=c_{\rm e}$ and ignore the second term 
in Eq.~(\ref{eq:dp2}), we find 
\begin{eqnarray}\label{eq:nodiss}
A_{\rm d} & = & \frac{1}{a}\left[2+
                \frac{\mathcal{H}^{\prime}}{\mathcal{H}^2} +3c_{\rm e}-6w_{\rm d}\right]\;, \nonumber \\
B_{\rm d} & = & \frac{1}{a^2}\left[3\left(c_{\rm e}-w_{\rm d}\right)(1+\frac{\mathcal{H}^{\prime}}{\mathcal{H}^2}-
                3w_{\rm d})+\frac{k^2 }{\mathcal{H}^2}c_{\rm e}-3a\frac{dw_{\rm d}}{da}\right]\;, \nonumber \\
S_{\rm d} & = & (1+w_{\rm d})\left[3\frac{d^{2}\phi}{da^{2}}+
                \frac{3}{a}\left(2+\frac{\mathcal{H}^{\prime}}{\mathcal{H}^2}-
                3w_{\rm d}\right)\frac{d\phi}{da} \right.\; \nonumber \\
          & - & \left.\frac{k^2}{a^2\mathcal{H}^2}\phi +\frac{3}{1+w_{\rm d}} \frac{d\phi}{da}\frac{dw_{\rm d}}{da}\right]\;, \nonumber
\end{eqnarray}
which coincide with the values in \cite{Abramo:2008ip} for $w_{\rm d}=const$ and 
$\frac{\mathcal{H}^{\prime}}{\mathcal{H}^2}=-
\frac{1}{2}$(matter dominated). 
We notice that for $w_{\rm d}=c_{\rm e}=c_{\rm a}=0$ the coefficients for matter density contrast are recovered.

\section{Poisson equation}\label{app:Poisson}
On sub-horizon scales, the basic equation describing the evolution of linear matter fluctuations is
\begin{equation}\label{dmm}
\ddot{\delta}_{\rm m}+2H(t)\dot{\delta}_{\rm m}+\frac{k^2}{a^2}\phi=0\;.
\end{equation}
In this context the Poisson equation in the Fourier space
is written as \citep{Lima:1996at}
\begin{equation}\label{eq:per-poiss-four}
k^2\phi=-4\pi Ga^2(\delta\rho+3\delta p)\;.
\end{equation}
where $\delta \rho=\delta \rho_{\rm m}+\delta \rho_{\rm d}$ and 
$\delta p=\delta p_{\rm m}+\delta p_{\rm d}$. Now using
$\delta p_{\rm m}=0$, $\delta p_{\rm d}=c_{\rm e}\delta \rho_{\rm d}$, 
$\delta \rho_{\rm m}=\rho_{\rm m}\delta_{\rm m}$, 
$\delta \rho_{\rm d}=\rho_{\rm d}\delta_{\rm d}$, 
and inserting the above quantities 
into Eq.(\ref{eq:per-poiss-four}), we arrive to
\begin{equation}\label{eq:pert-final1}
-\frac{k^2}{a^2}\phi=4\pi G [\rho_{\rm m}\delta_{\rm m}+(1+3c_{\rm e})
\rho_{\rm d}\delta_{\rm d}]\;,
\end{equation}
or
\begin{equation}\label{eq:pert-final}
-\frac{k^2}{a^2}\phi=\frac{3}{2}H^2[\Omega_{\rm m}\delta_{\rm m}+(1+3c_{\rm e})\Omega_{\rm d}\delta_{\rm d}]\;.
\end{equation}
Utilising the above equations it is easy to check that
\begin{equation}
\ddot{\delta}_{\rm m}+2H(t)\dot{\delta}_{\rm m}=
\frac{3H^{2}}{2}\left[\Omega_{\rm m}\delta_{\rm m}+\Omega_{\rm d}\delta_{\rm d}
(1+3c_{\rm e})\right]\;.
\end{equation}
Obviously for $c_{\rm e}=w_{\rm d}=const.$ the latter equation reduces to that of \cite{Abramo:2008ip} and \cite{
Mehrabi:2014ema}. 
Changing the variables from $t$ to $a$ we finally obtain Eq.~(\ref{odedelta}).

\bsp

\label{lastpage}

\end{document}